# Testing the topological nature of end states in antiferromagnetic atomic chains on superconductors

Lucas Schneider[1], Philip Beck[1], Levente Rózsa[2], Thore Posske[3,4], Jens Wiebe[1,*] and Roland Wiesendanger[1]

[1]Department of Physics, University of Hamburg, D-20355 Hamburg, Germany.

[2]Department of Physics, University of Konstanz, D-78457 Konstanz, Germany.

[3]I. Institute for Theoretical Physics, University of Hamburg, D-20355 Hamburg, Germany.

[4]Centre for Ultrafast Imaging, Luruper Chaussee 149, D-22761 Hamburg, Germany.

*E-mail: jwiebe@physnet.uni-hamburg.de

## Abstract

**Edge states forming at the boundaries of topologically non-trivial phases of matter are promising candidates for future device applications because of their stability against local perturbations. Magnetically ordered spin chains proximitized by an *s*-wave superconductor are predicted to enter a topologically non-trivial mini-gapped phase with zero-energy Majorana modes (MMs) localized at their ends. However, the presence of non-topological end states mimicking MM properties can spoil their unambiguous observation. Here, we report on a method to experimentally decide on the MM nature of end states observed for the first time in antiferromagnetic spin chains. Using scanning tunneling spectroscopy, we find end states at either finite or near-zero energy in Mn chains on Nb(110) or Ta(110), respectively, within a large minigap. By introducing a locally perturbing defect on one end of the chain, the end state on this side splits off from zero-energy while the one on the other side doesn't – ruling out their MM origin.  A minimal model shows that, while wide trivial minigaps hosting such conventional end states are easily achieved in antiferromagnetic spin chains, unrealistically large spin-orbit couplings are required to drive the system into the topologically nontrivial phase with MMs. The methodology of perturbing chains by local defects is a powerful tool to probe the stability of future candidate topological edge modes against local disorder.**

## Main

Hybrid systems of magnetic and superconducting materials in reduced dimensions have been of great interest in recent years, owing to the exciting emergent physics such as unconventional superconductivity and topological edge modes expected in these platforms[1–6]. In particular, there has recently been a focus on the interplay of antiferromagnetic materials proximity coupled to *s*-wave superconductors[7–11]. Since antiferromagnets possess no net magnetic moment in their magnetic unit cell, they do not have considerable stray fields which would destroy superconducting order. A coexistence of antiferromagnetism and superconductivity in, e.g., thin films on *s*-wave superconductors[8] or the Fe-based superconductors[12,13] was previously explained by the large size of Cooper pairs compared to the magnetic unit cell of the materials or by an unconventional $s_\pm$ type pairing symmetry[14]. The absence of the net magnetic moment also gives rise to an effective time-reversal symmetry (ETRS) in an antiferromagnet, consisting of physical time reversal inverting the spin directions and a spatial symmetry exchanging the antiferromagnetic sublattices.

In the limit of a single magnetic adatom or molecule on a superconducting surface, the interaction of its spin with the host material induces local quasiparticle states inside the superconducting gap known as Yu-Shiba-Rusinov (YSR) states[15–17]. When multiple of these impurities are close to each other, their YSR states split in energy as they start to couple[18,19]. Without spin-orbit coupling (SOC), such a splitting would be prohibited by the ETRS for a strictly antiferromagnetic alignment of the spins in adatom pairs. However, it has been shown recently that the splitting is allowed in the presence of SOC on a surface[19,20]. In larger arrays of magnetic impurities, the coupled YSR states form YSR sub-gap bands, which can potentially have non-trivial topology[3–6,21,22] and lead to the emergence of topologically protected Majorana modes (MMs) at the edges of the array[23–27]. In one dimension, a chain of magnetic impurities with topologically non-trivial YSR bands is expected to host zero-energy MMs at both ends for sufficient chain length. The Majorana number in spin chains can be interpreted as the parity of the number of spin-polarized bands crossing the Fermi level in the absence of superconducting pairing terms (c.f. Methods, Eq. (8)). Without SOC, the ETRS in antiferromagnetic chains leads to doubly degenerate excitations in the magnetic Brillouin zone. Thus, there is necessarily an even number of band crossings and a topologically trivial Majorana number. However, finite SOC breaks this symmetry and the degeneracies can be lifted. Thus, SOC or certain spatial symmetries theoretically open up possibilities for topologically non-trivial phases hosting MMs also in antiferromagnetic chains[9,10]. Experimental investigations of the low-energy electronic structure with a focus on such modes so far largely concentrated on ferromagnetic chains[24,26,28–33] and a few on spin-spirals[23,34], but studies of the antiferromagnetic case are sparse[33].

A general problem with the interpretation of such experimental data is the fact that near-zero-energy states can always appear as artifacts - e.g., induced by local defects or by a different electronic structure at the chain termination - in local tunneling spectroscopy measurements[32–34]. Therefore, a good understanding of the sample's underlying YSR band structure and its direct correlation with the observation of end states is clearly desired to pin down the nature of these end states. In this work, we additionally pursue a new strategy to test the MM nature of end states residing in a comparably large bulk minigap which we observed for scanning-tunnel-microscope-(STM)-tip-constructed[35] antiferromagnetic Mn chains on the atomically clean surfaces of Nb(110) and Ta(110): We intentionally locally perturb one end of the chain with a local defect. While MMs residing in a topological minigap which is wider in energy than the perturbation's energy scale are expected to remain either completely unaffected or will merely laterally shift, trivial end states will split in energy only on the perturbed side of the chain (see Supplementary Note 1). We compare the experimental findings to an effective single-particle model for an antiferromagnetic spin chain coupled to an $s$-wave superconductor.

## Antiferromagnetic Mn chains on Nb(110)

Single Mn atoms on clean Nb(110) and Ta(110) surfaces have been studied both experimentally[19,26,28,31,36,37] and theoretically[38–41], and offer a suitable platform for studying well-defined YSR arrays. In particular, it has been shown that the magnetic interaction between neighboring adatoms can be tuned from ferromagnetic to antiferromagnetic when varying the inter-atomic distance and the crystallographic direction connecting the atoms on the surface[19,36]. Spin-polarized measurements have revealed that densely packed linear chains of Mn atoms constructed along the $[1\bar{1}1]$ direction of the (110) surfaces (Fig. 1a) of Nb[36] and Ta (Supplementary Fig. 2) feature an out-of-plane antiferromagnetic ground state, in agreement with *ab-initio* calculations[40].

We start by presenting the results on antiferromagnetic Mn chains on Nb(110): Fig. 1b shows the topography of a $Mn_{40}$ chain together with examples of deconvoluted d$I$/d$V$ maps obtained at sub-gap energies ($\Delta_{Nb}$ = 1.51 meV, see Methods, Supplementary Note 3 and Supplementary Figure 3). A d$I$/d$V$ line profile along the same chain is presented in Fig. 1d. Additionally, the evolution of the sub-gap local density of states (LDOS) for $Mn_N$ chains on Nb(110) with increasing number of sites $N$ is shown in Fig. 1c, separately for the chain's left end (left panel), for the chain's bulk (central panel) and for the right end (right panel). There is a continuum of states visible in the energy range of 0.7 meV $< |E| <$ 1.5 meV, exhibiting standing wave patterns (c.f. Fig. 1d and the map at -1.04 meV in Fig. 1b) due to quasiparticle interference (QPI). This indicates the formation of dispersive YSR bands[26,31] which we analyze later on. In contrast to this, no bulk states are observed within a gapped region of $\pm\Delta_b$ = $\pm$0.7 meV for chain lengths exceeding $N$ = 6 sites. This is a surprising result since the maximal gaps previously found in ferromagnetic spin chains on superconducting surfaces were on the order of 50-180 μeV[23,26,29,31,42].

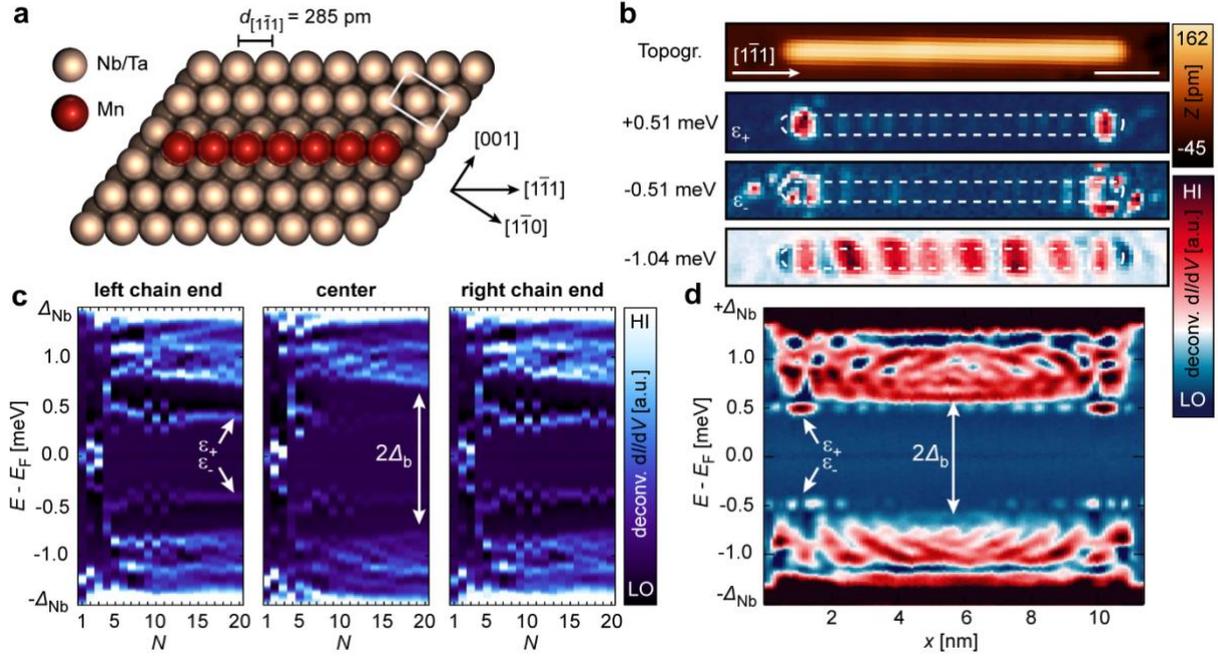

**Figure 1 | Geometry and low-energy electronic structure of antiferromagnetic Mn chains on Nb(110). a**, Sketch of Mn adatoms (red) positioned on neighboring hollow sites of the Nb(110) or Ta(110) lattice (beige) along the [1$\bar{1}$1] direction. **b**, Constant-current STM image (topography, top panel) of a Mn$_{40}$ chain and simultaneously acquired deconvoluted d$I$/d$V$ maps (bottom panels) at selected energies as indicated. The white scale bar corresponds to 2 nm. The apparent extent of the chain from the top panel is marked by white dashed boundaries. **c**, Sequence of deconvoluted d$I$/d$V$ spectra measured on the left end, in the center and on the right end of Mn$_N$ chains with increasing number of sites $N$. The emergence of the chain's bulk minigap $\Delta_b$ and of finite-energy end states $\varepsilon_{+/-}$ are indicated. Note that single atoms are added only to the right chain end during this measurement. **d**, Deconvoluted d$I$/d$V$ line-profile measured along the longitudinal axis through the center of the Mn$_{40}$ chain. The lateral position of the spectra is aligned with the topography in panel b. Parameters: $V_{stab}$ = -6 mV, $I_{stab}$ = 1 nA, $V_{mod}$ = 20 µV.

Inside the minigap $\Delta_b$, we find clearly localized end states at energies $\varepsilon_{+/-} = \pm 0.51$ meV (Fig. 1b). In addition to the end-state nature of these features, a small oscillatory component of their wave function is found to decay into the chain's bulk (see Figs. 1b,d). As it can be seen in Fig. 1c, these energetically isolated states form at energies $\varepsilon_{+/-}$ already for $N > 5$ and their energy is only faintly oscillating in energy for longer systems. This fast convergence of the end-state energy with increasing chain length agrees with the good localization of the features, i.e., interactions between the ends already vanish for short chains. Notably, the end states in the regime $8 < N < 14$ are energetically split into four eigenstates for odd $N$ while there are only two eigenstates for even $N$ (see Fig. 1c).

## Comparison to antiferromagnetic Mn chains on Ta(110)

In order to further investigate experimentally whether the appearance of such low-energy end states is a consequence of the antiferromagnetic spin alignment in the chain and thus not limited to only one experimental platform, we study structurally similar Mn chains on a clean Ta(110) surface ($\Delta_{Ta}$ = 0.64 meV). It has been shown previously that Mn atoms on a Ta(110) surface exhibit surprising similarities to Mn/Nb(110)[28,37] due to the identical number of valence electrons and the same crystal structure of the substrates, however, with a strongly enhanced SOC in Ta compared to Nb. Most notably, densely packed Mn chains along the [1$\bar{1}$1] direction are also found to be antiferromagnetically ordered (see Supplementary Fig. 2). Measurements of the low-energy electronic structure in these chains are presented in Fig. 2. The d$I$/d$V$ maps in Fig. 2a measured around a Mn$_{22}$ chain reveal the presence of well-localized end states with near-zero energy and a similar spatial appearance as the end states found in Mn chains on Nb(110), while higher-energy excitations are mostly localized in the chain's bulk. Furthermore, a weak oscillatory pattern can be seen in the map obtained at -0.28 meV, indicating the presence of QPI in the bulk states of this platform as well. The d$I$/d$V$ line-profile shown in Fig. 2b suggests that the bulk of the chain is electronically gapped by a minigap $\pm\Delta_b$ = $\pm$0.2 meV. Fig. 2c shows the length dependence of the LDOS features for chains with $10 < N < 22$. Here, it is also visible that end states of almost constant energy are present on both chain ends while the bulk remains gapped. On the left, unperturbed end, a damped even-odd oscillation in the end-state energies is observed again, as for the Nb case. The overall appearance of the low-energy electronic structure is very similar to Mn/Nb(110). It is therefore natural to conjecture that the end states in the Mn/Ta(110) platform have a similar origin as in the Mn/Nb(110) case. However, since the end states in Mn/Ta(110) are very close to or at zero energy, the question arises whether or not this system realizes a topological superconductor with the accompanying near-zero energy MMs at its ends.

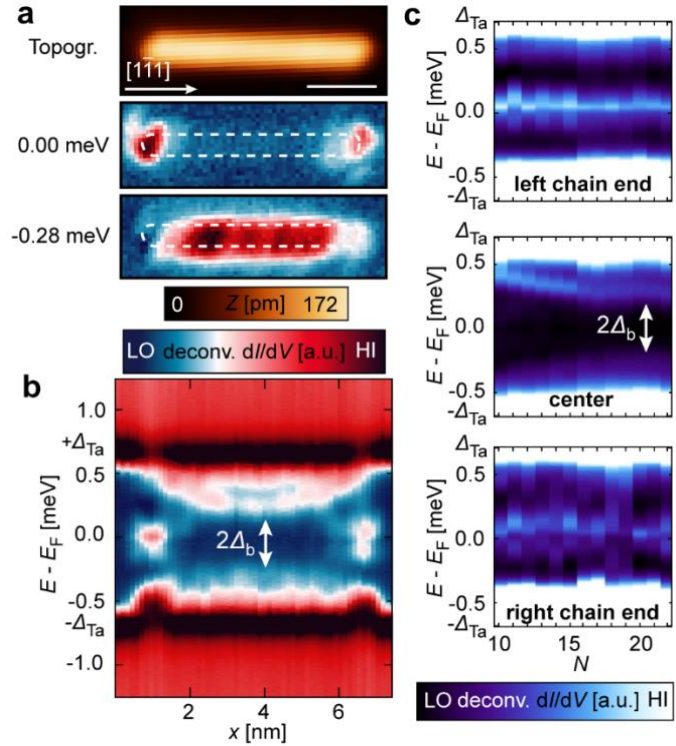

**Figure 2 | Sub-gap electronic structure of antiferromagnetic Mn chains on Ta(110). a**, Constant-current STM image (topography, top panel) of a $Mn_{22}$ chain and simultaneously acquired deconvoluted d$I$/d$V$ maps (bottom panels) at selected energies as indicated. The white scale bar corresponds to 2 nm. The apparent extent of the chain from the top panel is marked by white dashed boundaries. **b**, Deconvoluted d$I$/d$V$ line-profile measured along the longitudinal axis through the center of the $Mn_{22}$ chain. The lateral position of the spectra is aligned with the topography in panel a. The arrow indicates the chain's bulk minigap. **c**, Sequence of deconvoluted d$I$/d$V$ spectra measured at the left end, in the center and at the right end of $Mn_N$ chains with increasing number of sites $N$. The emergence of a bulk minigap $\Delta_b$ (marked) and of end states with energies close to the Fermi energy $E_F$ can be observed. Parameters: $V_{stab}$ = -2.5 mV, $I_{stab}$ = 1 nA, $V_{mod}$ = 20 µV.

## Perturbation of the end states by local defects

For testing the topologically non-trivial or trivial nature of the end states, the influence of local defects on the end states in Mn/Nb(110) and Mn/Ta(110) chains is studied. These defects can be either of magnetic or of non-magnetic origin. It has been shown previously that the energy of individual YSR states is very sensitive to variations in adsorption geometries[43,44], defects like local oxygen impurities[45], or local charge density[46]. Also hydrogenation of adatoms has been shown to drastically alter their magnetic properties and, importantly, their exchange coupling to the substrate[47]. Since the exchange coupling strength is one of the main factors determining the YSR state energies of magnetic impurities[17], hydrogenated Mn atoms at the end of the chain are expected to have clearly shifted YSR state energies compared to unperturbed Mn atoms. Fig. 3a shows an example of a $Mn_{20}$ chain on Nb(110) with a dark spot visible on the left side of the chain, which is presumably adsorbed hydrogen or another weakly bound surface adsorbate trapped at an oxygen defect of the Nb(110) surface. When measuring d$I$/d$V$ maps at sub-gap energies on this Mn chain, the left and the right end state have slightly different energies (±0.45 meV and ±0.39 meV, respectively). When removing the defect next to the chain by local voltage pulses (Fig. 3b), both end states appear at the same energy (±0.35 meV) again. A similar effect can actually be seen in the data of Fig. 1c already where the end state on the right end is found to oscillate in energy for 15 < $N$ < 20 while the state on the left end remains at fixed energy. This unambiguously proves the local nature of these states, in clear contrast to non-local, spatially correlated states like MMs or their precursors[26] (see Supplementary Note 1 for the effect of potential disorder within a minimal model for antiferromagnetic YSR chains).

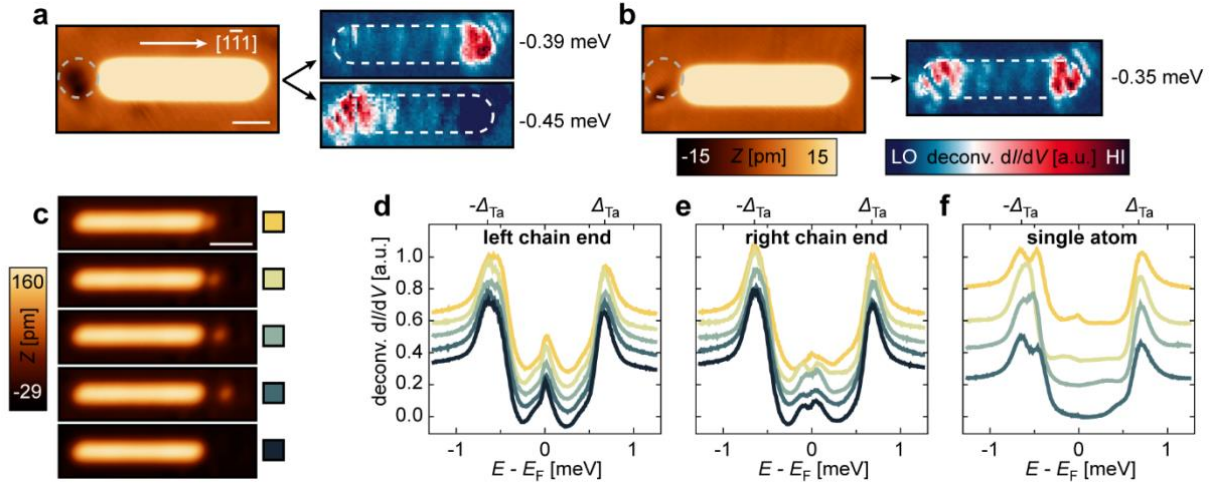

**Figure 3 | Response of the end state energies to local defects. a**, Constant-current image of a Mn$_{20}$ chain on Nb(110) (left panel) with adjusted contrast to highlight the defect on the left side of the image which is marked by the gray dashed circle. The white bar corresponds to 1 nm. Deconvoluted d$I$/d$V$ maps at selected energies (right panels) reveal that the two end states have different energies. The apparent extent of the chain from the left panel is marked by white dashed boundaries. **b**, Constant-current image of the same Mn$_{20}$ chain after applying bias voltage pulses to remove the local defect. A deconvoluted d$I$/d$V$ map at a selected energy (right) shows that the end states now have equal energy. Parameters: $V_{stab}$ = -6 mV, $I_{stab}$ = 1 nA, $V_{mod}$ = 20 μV. **c**, Constant-current STM images of a Mn$_{22}$ chain on Ta(110) and a single Mn atom which is subsequently moved to positions with different distances from the right chain end. The bottommost panel shows the unperturbed chain. The white bar corresponds to 2 nm. Parameters: $V_{stab}$ = -20 mV, $I_{stab}$ = 0.2 nA. **d**, Deconvoluted d$I$/d$V$ spectra measured on the left chain end, **e**, on the right chain end and **f**, on the single Mn atom for the different adatom distances depicted in panel c. The spectra are vertically offset for clarity. Their color indicates which panel in c they belong to. Parameters: $V_{stab}$ = -2.5 mV, $I_{stab}$ = 1 nA, $V_{mod}$ = 20 μV.

We performed a similar experiment for Mn chains on Ta(110) using a single Mn atom as the local defect. Fig. 3c shows topographies of a Mn$_{22}$ chain with an additional Mn atom, whose distance to the right chain end is varied. d$I$/d$V$ spectroscopy on the left and right chain end as well as on the single atom reveals the coupling between both structures (Fig. 3d-f): as the single atom approaches the chain, its sub-gap spectral character is altered. The same holds for the right chain end, which directly interacts with the single atom because of their close proximity, i.e. the peak position of the end state slightly shifts. In contrast, the left chain end is not altered, proving that the end states are entirely local instead of collective properties of the chain. This is not expected for MMs or their precursors, where moving a magnetic atom close to one chain end merely laterally shifts the spatial location of the zero-energy end state on that side, or shifts the energy of the lowest-lying state simultaneously at both ends of the chain[26]. The observed energetical splitting of the two end states with respect to each other indicates that in the undisturbed chains, two states are localized, one at each end, which are degenerate due to the antiferromagnetic spin structure, in contrast to MMs or their precursors that form a single fermionic state with enhanced intensities at both ends.

**Minimal model for antiferromagnetic YSR chains**

To understand the nature of these trivial end states, we construct a minimal theoretical model for antiferromagnetic YSR chains. A single-particle model following Refs.[3,4,6,21,22] successfully describes the sub-gap electronic bands in ferromagnetic YSR chains, especially when extended with local potential scattering[26,31]. We extend these models by studying an antiferromagnetic chain on a superconducting substrate including Rashba SOC and arrive at the following minimal next-nearest-neighbor model Hamiltonian (see Methods for details):

$$\mathcal{H} = -E_0 \sum_{i=1}^{N} c_i^\dagger c_i - t_1 \sum_{i=1}^{N-1} (c_i^\dagger c_{i+1} + \text{h.c.}) - t_2 \sum_{i=1}^{N-2} (c_i^\dagger c_{i+2} + \text{h.c.})$$
$$- \Delta_1 \sum_{i=1}^{N-1} (c_i^\dagger c_{i+1}^\dagger + \text{h.c.}) - \Delta_2 \sum_{i=1}^{N-2} (c_i^\dagger c_{i+2}^\dagger + \text{h.c.}). \quad (1)$$

Here, $c_i^\dagger$, $c_i$ represent the creation and annihilation operators of YSR states at site $i$ of a one-dimensional chain with $N$ sites. The on-site energies $-E_0$ would correspond to the YSR state energies of the individual Mn impurity. The model includes nearest-neighbor (NN: index 1) and next-nearest-neighbor (NNN: index 2) hopping ($t_1$, $t_2$) and superconducting pairing ($\Delta_1$, $\Delta_2$) (Fig. 4e). The model connects to the Kitaev chain[48] if only the NN terms are kept.

Notably, a perfectly collinear antiferromagnetic ordering is characterized by an ETRS (see Methods) consisting of physical time reversal inverting the spin directions and a translation by the distance between the chain atoms[10]. In the present Hamiltonian expressed in the basis of atomic YSR states, the ETRS implies that hopping terms are only allowed between atomic YSR states with the same spin, i.e. between NNNs in the present model (see Methods and Fig. 4e). In contrast, the effective superconducting pairing $\Delta_1$ can easily be induced between adjacent atoms, since their spins are anti-aligned, and will be suppressed for NNNs ($\Delta_2$), where the spins have a parallel alignment. For non-zero SOC these restrictions are lifted (see Methods), and the ratio of the coefficients induced by the SOC and present without the SOC may be estimated by the dimensionless parameter $\alpha_R/\hbar v_F$, where $\alpha_R$ is the Rashba parameter and $v_F$ is the Fermi velocity. This parameter was estimated to be $k_h/k_{F,0} = 0.094$ for Mn/Nb(110) in Ref. [26]. We consequently pay particular attention to the case $t_1 \ll t_2$, and $\Delta_1 \gg \Delta_2$ of our minimal model and further assume $E_0 \approx 0.0$ meV, which is motivated by the experimental YSR state energies for $N$ = 1,2,3 (Fig. 1c).

Experimentally, information about the sub-gap bands' dispersion can be extracted from one-dimensional quasiparticle interference (QPI) measurements[26,31] (see Supplementary Note 4 and the extracted dispersion of the scattering vectors in Fig. 4c). We adjust the parameters of the minimal model such that the dispersions of the possible scattering vectors $q_1$, $q_2$ in Fig. 4b extracted from the calculated dispersion in Fig. 4a reasonably fit the experimental dispersions of Fig. 4c. Using these parameters, we calculate the spatially resolved LDOS for various chain lengths $N$ (Fig. 4d, Supplementary Fig. 4, and Methods) which, given the few free parameters, agree surprisingly well with the experimental data in Fig. 1c,d. In particular, the minimal model nicely reproduces the appearance of non-zero energy end states with an oscillatory decay towards the chain center at energies $\varepsilon_{+/-}$ in a large minigap $\Delta_b$, the fast convergence of these end states towards $\varepsilon_{+/-}$ already for short chains (Supplementary Fig. 4), as well as the dispersive nature of the bulk states.

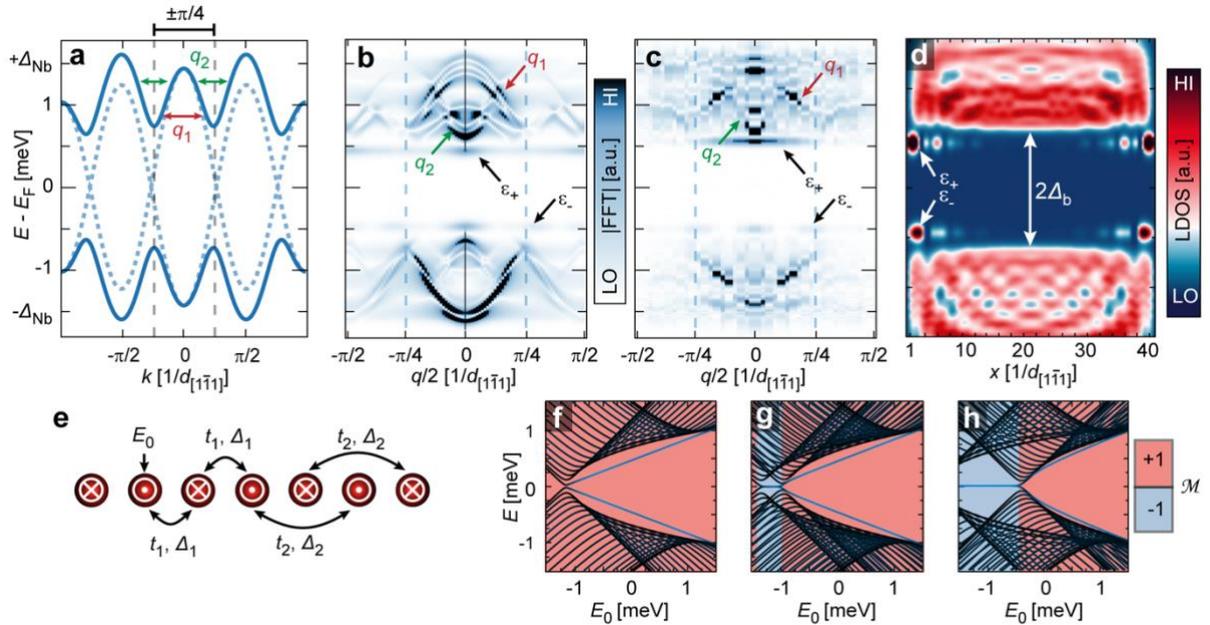

**Figure 4 | Minimal model for antiferromagnetic YSR chains and experimental QPI results. a,** YSR band structure from the minimal model using the parameters $E_0 = 0.0$ meV, $t_1 = 0.1$ meV, $t_2 = 0.6$ meV, $\Delta_1 = 0.0$ meV (dashed lines) as well as $\Delta_1 = 0.5$ meV (solid lines), and $\Delta_2 = 0.0$ meV. The position of $k = \pm\pi/4$ is marked by dashed lines. **b,** Absolute values of the line-wise FFT of a calculated LDOS line-profile similar panel d using $N$ = 100 sites. **c,** Absolute values of the line-wise FFT of the experimental d$I$/d$V$ line-profile in Fig. 1d. The supposed dispersive scattering vectors $q_1$, $q_2$ and the edge state energies $\varepsilon_{+,-}$ are marked in panels a-c. **d,** Calculated LDOS along a chain of $N$ = 40 sites using the parameters from panels a and b. The energies of the finite-energy end states $\varepsilon_{+/-}$ and the chain's bulk minigap $\Delta_b$ are marked. **e,** Sketch of the parameters in an antiferromagnetically ordered chain considered in the minimal model of the YSR states. The symbols indicate up (⊙) and

down (⊗) oriented magnetic moments, respectively. $t_1$ ($t_2$) is the NN (NNN) hopping, $\Delta_1$ ($\Delta_2$) is the NN (NNN) pairing term and $E_0$ is the on-site potential energy (see Methods). **f**, Eigenenergies of a finite-size chain with *N* = 40 for varying on-site energy $E_0$ and hopping parameters $t_1 = 0.0$ meV, $t_2 = 0.6$ meV. **g**, Same as panel e, but with a small ETRS breaking term, $t_1 = 0.1$ meV, $t_2 = 0.6$ meV. **h**, Same as panel e and g, but with an unrealistically large ETRS breaking term, $t_1 = 0.4$ meV, $t_2 = 0.6$ meV. $\Delta_1 = 0.5$ meV, $\Delta_2 = 0.0$ meV are chosen for all panels f-h. The Majorana number $\mathcal{M}$ of the system is trivial (+1) in the red regions and non-trivial (-1) in the blue regions (see Methods for details). The dark blue lines in panels f-h are end states of the system.

Having substantiated the minimal model by comparison to the experimental data, we investigate its parameter-dependent phases and the topological nature of the edge states in the following. Figs. 4f-h show the energy spectrum of a 40-site chain vs. the on-site energy $E_0$ for different values of the hopping terms $t_1$ and $t_2$. It can be seen that the topologically non-trivial phase of the Kitaev chain is entirely quenched for $t_1 \to 0$ which corresponds to zero SOC (Fig. 4f). This is a consequence of the ETRS, which results in a Kramers degeneracy of all states, hence to a topologically trivial Majorana number. This finding agrees with previous studies, showing that topologically non-trivial phases cannot be found in antiferromagnetic YSR chains without either strong SOC or additional supercurrents[9,10] breaking the ETRS. However, it is found that the end states at finite energy are formed even in this topologically trivial regime for $t_2 > t_1$. Each of these end states is twofold degenerate for even-length chains and shows a small splitting for odd-length chains where the ETRS is broken (cf. the observed experimental splittings in the regime 8 < *N* < 14 for odd *N* in Fig. 1c); the degeneracy is restored for semi-infinite chains ($N \to \infty$). The fact that they merge with the continuum of states in Figs. 4f-h for large $E_0$ without a gap closure further proves their topologically trivial origin.

For non-zero values of $t_1$, corresponding to finite SOC, the non-zero-energy end states in some part of the topologically non-trivial phase are preserved (Fig. 4g). But now, also a topologically non-trivial phase with zero-energy MMs is recovered, which only takes up a large proportion of the phase space for unrealistically large values of SOC (Fig. 4h).

## Conclusions & Outlook

In conclusion, we have shown that the in-gap quasiparticle structure of dense, antiferromagnetic YSR chains can be qualitatively described by a minimal model. In contrast to ferromagnetic chains, the antiferromagnetic structure facilitates the formation of a large minigap in the YSR band even without SOC. This observation may be qualitatively explained by the fact that all YSR states in a ferromagnetic chain possess the same spin polarization, meaning that the *s*-wave superconducting pairing in the substrate may only open a gap in the spectrum for non-zero SOC. For the antiferromagnetic chain the localized YSR states at the sites have an alternating spin polarization, which enables pairing between these states without SOC and can lead to a considerably larger minigap. This large size of the minigap naturally leads to a strong localization of potential end states, since their extension is inversely proportional to the minigap width[26,49]. Therefore, antiferromagnetic chains appear to be more suitable for the realization of well-localized end states than their ferromagnetic counterparts. However, as visible from our simulations in Fig. 4f and as was shown in Refs. [9] and [10], the formation of the topologically non-trivial phase now is a threshold effect: the SOC has to compete with the pairing potential, meaning that unrealistically large values of SOC have to be assumed in order to enable the formation of MMs. This is in contrast to the ferromagnetic chain, which is gapless in the absence of SOC and consequently an arbitrarily weak SOC may drive it into the topologically non-trivial regime. The most promising path towards combining the large minigap of antiferromagnetic chains with the easily achievable topologically non-trivial phases of ferromagnetic chains lies in the realization of YSR bands formed by non-collinear spin configurations[23,50].

Our minimal model furthermore reproduces the presence of intriguing finite-energy end states in the antiferromagnetic chains which are not topologically protected and whose energy can be tuned by local potentials. A corresponding topologically trivial phase has been characterized by Pientka *et al*. as a two-channel *p*-wave superconducting wire where the interaction between two pairs of MMs lifts their energy to finite values and destroys their topological protection[3,51]. However, we want to emphasize that this separation into two channels is not straightforward if SOC is present, and thus, this argument must not be taken at face value. Coincidentally, the trivial end states may appear at near-zero energy by local potentials where they could be misinterpreted as MMs. We have shown here that a local perturbation of the end states with defects is a distinct way to prove their topologically trivial or non-trivial nature. This methodology can be used on other sample systems to probe the stability of candidate topological edge modes against local disorder.

# References

1. Nadj-Perge, S., Drozdov, I. K., Bernevig, B. A. & Yazdani, A. Proposal for realizing Majorana fermions in chains of magnetic atoms on a superconductor. *Phys. Rev. B* **88**, 020407 (2013).
2. Li, J. *et al.* Topological superconductivity induced by ferromagnetic metal chains. *Phys. Rev. B* **90**, 235433 (2014).
3. Pientka, F., Glazman, L. I. & von Oppen, F. Topological superconducting phase in helical Shiba chains. *Phys. Rev. B* **88**, 155420 (2013).
4. Li, J. *et al.* Two-dimensional chiral topological superconductivity in Shiba lattices. *Nat. Commun.* **7**, 12297 (2016).
5. Schecter, M., Flensberg, K., Christensen, M. H., Andersen, B. M. & Paaske, J. Self-organized topological superconductivity in a Yu-Shiba-Rusinov chain. *Phys. Rev. B* **93**, 140503 (2016).
6. Brydon, P. M. R., Das Sarma, S., Hui, H.-Y. & Sau, J. D. Topological Yu-Shiba-Rusinov chain from spin-orbit coupling. *Phys. Rev. B* **91**, 064505 (2015).
7. Díaz, S. A., Klinovaja, J., Loss, D. & Hoffman, S. Majorana bound states induced by antiferromagnetic skyrmion textures. *Phys. Rev. B* **104**, 214501 (2021).
8. Lo Conte, R. *et al.* Coexistence of antiferromagnetism and superconductivity in Mn/Nb(110). *Phys. Rev. B* **105**, L100406 (2022).
9. Heimes, A., Kotetes, P. & Schön, G. Majorana fermions from Shiba states in an antiferromagnetic chain on top of a superconductor. *Phys. Rev. B* **90**, 060507 (2014).
10. Heimes, A., Mendler, D. & Kotetes, P. Interplay of topological phases in magnetic adatom-chains on top of a Rashba superconducting surface. *New J. Phys.* **17**, 023051 (2015).
11. Kobiałka, A., Sedlmayr, N. & Ptok, A. Majorana bound states in a superconducting Rashba nanowire in the presence of antiferromagnetic order. *Phys. Rev. B* **103**, 125110 (2021).
12. Manna, S. *et al.* Interfacial superconductivity in a bi-collinear antiferromagnetically ordered FeTe monolayer on a topological insulator. *Nat. Commun.* **8**, 14074 (2017).
13. Aluru, R. *et al.* Atomic-scale coexistence of short-range magnetic order and superconductivity in $Fe_{1+y}Se_{0.1}Te_{0.9}$. *Phys. Rev. Mater.* **3**, 084805 (2019).
14. Fernandes, R. M. *et al.* Unconventional pairing in the iron arsenide superconductors. *Phys. Rev. B* **81**, 140501 (2010).
15. Heinrich, B. W., Pascual, J. I. & Franke, K. J. Single magnetic adsorbates on s -wave superconductors. *Prog. Surf. Sci.* **93**, 1–19 (2018).
16. Yazdani, A., Eigler, D. M., Lutz, C. P., Jones, B. A. & Crommie, M. F. Probing the Local Effects of Magnetic Impurities on Superconductivity. *Science* **275**, 1767–1770 (1997).
17. Franke, K. J., Schulze, G. & Pascual, J. I. Competition of Superconducting Phenomena and Kondo Screening at the Nanoscale. *Science* **332**, 940–944 (2011).
18. Ruby, M., Heinrich, B. W., Peng, Y., Von Oppen, F. & Franke, K. J. Wave-Function Hybridization in Yu-Shiba-Rusinov Dimers. *Phys. Rev. Lett.* **120**, 156803 (2018).
19. Beck, P. *et al.* Spin-orbit coupling induced splitting of Yu-Shiba-Rusinov states in antiferromagnetic dimers. *Nat. Commun.* **12**, 2040 (2021).
20. Kim, Y., Zhang, J., Rossi, E. & Lutchyn, R. M. Impurity-Induced Bound States in Superconductors with Spin-Orbit Coupling. *Phys. Rev. Lett.* **114**, 236804 (2015).
21. Röntynen, J. & Ojanen, T. Topological Superconductivity and High Chern Numbers in 2D Ferromagnetic Shiba Lattices. *Phys. Rev. Lett.* **114**, 236803 (2015).
22. Röntynen, J. & Ojanen, T. Chern mosaic: Topology of chiral superconductivity on ferromagnetic adatom lattices. *Phys. Rev. B* **93**, 094521 (2016).
23. Kim, H. *et al.* Toward tailoring Majorana bound states in artificially constructed magnetic atom chains on elemental superconductors. *Sci. Adv.* **4**, eaar5251 (2018).
24. Nadj-Perge, S. *et al.* Observation of Majorana fermions in ferromagnetic atomic chains on a superconductor. *Science* **346**, 602–607 (2014).
25. Kezilebieke, S. *et al.* Topological superconductivity in a van der Waals heterostructure. *Nature* **588**, 424–428 (2020).
26. Schneider, L. *et al.* Precursors of Majorana modes and their length-dependent energy oscillations probed at both ends of atomic Shiba chains. *Nat. Nanotechnol.* **17**, 384–389 (2022).
27. Palacio-Morales, A. *et al.* Atomic-scale interface engineering of Majorana edge modes in a 2D magnet-superconductor hybrid system. *Sci. Adv.* **5**, eaav6600 (2019).
28. Beck, P., Schneider, L., Wiesendanger, R. & Wiebe, J. Effect of substrate spin-orbit coupling on the

29. Ruby, M. *et al.* End States and Subgap Structure in Proximity-Coupled Chains of Magnetic Adatoms. *Phys. Rev. Lett.* **115**, 197204 (2015).
30. Ruby, M., Heinrich, B. W., Peng, Y., von Oppen, F. & Franke, K. J. Exploring a Proximity-Coupled Co Chain on Pb(110) as a Possible Majorana Platform. *Nano Lett.* **17**, 4473–4477 (2017).
31. Schneider, L. *et al.* Topological Shiba bands in artificial spin chains on superconductors. *Nat. Phys.* **17**, 943–948 (2021).
32. Liebhaber, E. *et al.* Quantum spins and hybridization in artificially-constructed chains of magnetic adatoms on a superconductor. *Nat. Commun.* **13**, 2160 (2022).
33. Küster, F. *et al.* Non-Majorana modes in diluted spin chains proximitized to a superconductor. *Proc. Natl. Acad. Sci.* **119**, 2112.05708 (2022).
34. Schneider, L. *et al.* Controlling in-gap end states by linking nonmagnetic atoms and artificially-constructed spin chains on superconductors. *Nat. Commun.* **11**, 4707 (2020).
35. Eigler, D. M. & Schweizer, E. K. Positioning single atoms with a scanning tunnelling microscope. *Nature* **344**, 524–526 (1990).
36. Schneider, L., Beck, P., Wiebe, J. & Wiesendanger, R. Atomic-scale spin-polarization maps using functionalized superconducting probes. *Sci. Adv.* **7**, eabd7302 (2021).
37. Beck, P., Schneider, L., Wiesendanger, R. & Wiebe, J. Systematic study of Mn atoms, artificial dimers and chains on superconducting Ta(110). *arXiv* 2205.10073 (2022).
38. Nyári, B. *et al.* Relativistic first-principles theory of Yu-Shiba-Rusinov states applied to Mn adatoms and Mn dimers on Nb(110). *Phys. Rev. B* **104**, 235426 (2021).
39. Crawford, D. *et al.* Majorana modes with side features in magnet-superconductor hybrid systems. *npj Quantum Mater.* **7**, 117 (2022).
40. Lászlóffy, A., Palotás, K., Rózsa, L. & Szunyogh, L. Electronic and Magnetic Properties of Building Blocks of Mn and Fe Atomic Chains on Nb(110). *Nanomaterials* **11**, 1933 (2021).
41. Crawford, D. *et al.* Increased localization of Majorana modes in antiferromagnetic chains on superconductors. *arXiv* 2210.11587 (2022).
42. Feldman, B. E. *et al.* High-resolution studies of the Majorana atomic chain platform. *Nat. Phys.* **13**, 286–291 (2017).
43. Schneider, L. *et al.* Magnetism and in-gap states of 3d transition metal atoms on superconducting Re. *npj Quantum Mater.* **4**, 42 (2019).
44. Ruby, M., Peng, Y., von Oppen, F., Heinrich, B. W. & Franke, K. J. Orbital Picture of Yu-Shiba-Rusinov Multiplets. *Phys. Rev. Lett.* **117**, 186801 (2016).
45. Odobesko, A. *et al.* Observation of tunable single-atom Yu-Shiba-Rusinov states. *Phys. Rev. B* **102**, 174504 (2020).
46. Liebhaber, E. *et al.* Yu–Shiba–Rusinov States in the Charge-Density Modulated Superconductor NbSe 2. *Nano Lett.* **20**, 339–344 (2020).
47. Khajetoorians, A. A. *et al.* Tuning emergent magnetism in a Hund's impurity. *Nat. Nanotechnol.* **10**, 958–964 (2015).
48. Kitaev, A. Y. Unpaired Majorana fermions in quantum wires. *Physics-Uspekhi* **44**, 131–136 (2001).
49. Peng, Y., Pientka, F., Glazman, L. I. & von Oppen, F. Strong Localization of Majorana End States in Chains of Magnetic Adatoms. *Phys. Rev. Lett.* **114**, 106801 (2015).
50. Klinovaja, J., Stano, P., Yazdani, A. & Loss, D. Topological Superconductivity and Majorana Fermions in RKKY Systems. *Phys. Rev. Lett.* **111**, 186805 (2013).

## Methods

**Experimental procedures**

The experiments were performed in a home-built STM setup under ultra-high-vacuum at a base temperature of $T$ = 320 mK[52]. Nb(110) and Ta(110) single crystals were used as a substrate and cleaned by high-temperature flashes to $T$ > 2700 K with an e-beam heater. In this way, atomically clean surfaces with only few residual oxygen impurities on the surface can be obtained for both materials, as shown previously[37,53]. Subsequently, single Mn atoms were deposited onto the surface of the clean substrates held at low temperatures ($T$ < 7 K), resulting in a statistical distribution of adatoms. We use superconducting Nb tips made from mechanically cut and sharpened high-purity Nb wire. The tips were flashed in situ to about 1500 K to remove residual contaminants. STM images were obtained by regulating the tunneling current $I_{stab}$ to a constant value with a feedback loop while applying a constant bias voltage $V_{stab}$ across the tunneling junction. For measurements of differential tunneling conductance (d$I$/d$V$) spectra, the tip was stabilized at bias voltage $V_{stab}$ and current $I_{stab}$ as individually noted in the figure captions. In a next step, the feedback loop was switched off and the bias voltage was swept from -$V_{stab}$ to +$V_{stab}$. The d$I$/d$V$ signal was measured using standard lock-in techniques with a small modulation voltage $V_{mod}$ (RMS) of frequency $f$ = 4.142 kHz added to $V_{stab}$. d$I$/d$V$ line-profiles and maps were acquired recording multiple d$I$/d$V$ spectra along a one-dimensional line or a two-dimensional grid of lateral positions on the sample, respectively. Note that we chose stabilization parameters at which the contribution of Andreev reflections and direct Cooper pair tunneling can be neglected (see Supplementary Note 3). The use of superconducting Nb tips increases the effective energy resolution of the experiment beyond the Fermi-Dirac limit[54]. However, the differential tunneling conductance d$I$/d$V$ measured with superconducting tips is proportional to the convolution of the sample's local density of states (LDOS) and the superconducting tip density of states (DOS). Consequently, STS data need to be numerically deconvoluted in order to resemble the sample's LDOS, as it is typically known for the interpretation of STS experiments. After careful deconvolution of the spectra, the superconducting gaps of the Nb and Ta surfaces are found to be $\Delta_{Nb}$ = 1.51 meV and $\Delta_{Ta}$ = 0.64 meV, respectively (see Refs. [26,31] for Nb and Supplementary Fig. 3 for Ta). We show only deconvoluted data throughout the manuscript (see Supplementary Note 3 for details). Mn chains were constructed using lateral atom manipulation[36,37] techniques at low tunneling resistances of $R \approx$ 30 - 60 k$\Omega$.

**Minimal model for YSR bands in antiferromagnetic chains**

Pientka *et al.* showed in Ref. [3] that the low-energy electronic structure of a single orbital chain of classical magnetic moments with a helical spin texture embedded in a three-dimensional superconducting host can be reduced to an effective Bogoliubov-de-Gennes Hamiltonian on a basis of projected YSR states. Subsequent models of the same type included Rashba-type SOC in ferromagnetic chains in Refs. [2,4,6] and non-zero potential scattering and particle-hole asymmetric spectral weights in Refs. [31,55]. Here, we combine the SOC with non-zero potential scattering and a general spin structure, where the impurity atoms are assumed to be identical apart from the orientation of their classical spin. The effective Hamiltonian describing the YSR subgap band is given by

$$\mathcal{H} = \frac{1}{2}\sum_{i,j}\begin{bmatrix}\tilde{c}_i^\dagger & \tilde{c}_i\end{bmatrix}\begin{bmatrix}h_{ij} & \Delta_{ij} \\ \Delta_{ij}^\dagger & -h_{ij}^*\end{bmatrix}\begin{bmatrix}\tilde{c}_j \\ \tilde{c}_j^\dagger\end{bmatrix}, \quad (2)$$

with the matrix elements expressed as

$$h_{ij} = -E_0\delta_{ij} + h_{ij}^{(0)}\langle\uparrow(i)|\uparrow(j)\rangle + h_{ij}^{(\alpha)}\langle\uparrow(i)|i\sigma^y|\uparrow(j)\rangle, \quad (3)$$

$$\Delta_{ij} = \Delta_{ij}^{(0)}\langle\uparrow(i)|\downarrow(j)\rangle + \Delta_{ij}^{(\alpha)}\langle\uparrow(i)|i\sigma^y|\downarrow(j)\rangle, \quad (4)$$

where $E_0$ is the single-impurity YSR energy. The real-valued coefficients $h_{ij}^{(0)}, h_{ij}^{(\alpha)}, \Delta_{ij}^{(0)}$ and $\Delta_{ij}^{(\alpha)}$ are material-specific constants, decaying with the distance $r = |\mathbf{r}_i - \mathbf{r}_j|$ as $\propto r^{-1}e^{ik_F r - r/\xi_0}$ for an isotropic electronic structure in three dimensions. Crucially, these terms do not depend on the magnetic structure of the chain but only on the electronic structure of the substrate and on the magnetic and non-magnetic scattering amplitudes of the impurities. Constants with superscript $(0)$ remain finite at zero SOC, while the terms with superscript $(\alpha)$ vanish. Exemplary formulae for these coefficients are given in Refs. [4,6]. We refrain from giving their full form here since they are not used explicitly in this manuscript, where the coefficients are fit to the experimentally observed YSR band structure, remaining consistent with the above form. Following a similar analysis to Ref. [4], the most important conclusion is that the $(\alpha)$ terms are linear in the dimensionless SOC parameter $\alpha_R/\hbar v_F$ in leading order, making them typically two orders of magnitude smaller than the $(0)$ terms. Due to the oscillatory decay of the parameters with the distance, it might be possible but considerably difficult to design a system where the $(\alpha)$ and $(0)$ terms have comparable magnitude; decreasing the Fermi velocity in flat bands may also provide a way for achieving this.

The magnetic structure only enters in the matrix elements of the vectors $|\uparrow(i)\rangle = (\cos\vartheta_i/2\, e^{-i\varphi_i/2}, \sin\vartheta_i/2\, e^{i\varphi_i/2})$ and $|\downarrow(i)\rangle = (\sin\vartheta_i/2\, e^{-i\varphi_i/2}, -\cos\vartheta_i/2\, e^{i\varphi_i/2})$, which are eigenvectors of the spin operator $\mathbf{S}_i\boldsymbol{\sigma}$ with $\mathbf{S}_i = (\sin\vartheta_i\cos\varphi_i, \sin\vartheta_i\sin\varphi_i, \cos\vartheta_i)$ describing the magnetization direction of the $i$th impurity. In Eqs. (3) and (4), the Pauli matrix $\sigma^y$ enters due to the Rashba term when assuming that the chain is along the $x$ direction; for a different chain direction or symmetry class, a different spin direction would be selected by the SOC. In the selected representation, the matrices possess the symmetry $\mathbf{h} = \mathbf{h}^\dagger$ and $\mathbf{\Delta} = -\mathbf{\Delta}^T$, and the particle-hole constraint may be represented in the usual form as $C = \tau^x K$, where $\tau^x$ exchanges the creation and annihilation operators and $K$ denotes complex conjugation.

We assume an antiferromagnetic spin structure with alternating sublattices $A$ and $B$ with $\vartheta_A = \pi - \vartheta_B = \vartheta$ and $\varphi_A = \varphi_B + \pi = \varphi$. This implies $\langle \uparrow(A)|\uparrow(B)\rangle = 0$ and $\langle \uparrow(A)|\downarrow(A)\rangle = 0$, i.e., the hopping $h_{ij}$ vanishes between sites at an odd distance and the pairing $\Delta_{ij}$ vanishes between sites at an even distance, as discussed in the main text. Due to the alternating structure, we restrict the Hamiltonian to NN and NNN terms, which we justify by the compact spatial structure of the YSR states of single Mn atoms on Nb(110) and on Ta(110) and their comparably weak coupling for interatomic distances above 1 nm[19,28,31,37]. The matrices read

$$\boldsymbol{h} = \begin{matrix} A \\ B \end{matrix} \begin{bmatrix} \ddots & & & & & & \\ \dots & t_2 - it'_2 s_\parallel & t_1 s_\perp & -E_0 & -t_1 s_\perp & t_2 + it'_2 s_\parallel & \dots \\ \dots & t_2 + it'_2 s_\parallel & -t_1 s_\perp^* & -E_0 & t_1 s_\perp^* & t_2 - it'_2 s_\parallel & \dots \\ & & & & & & \ddots \end{bmatrix}, \quad (5)$$

$$\boldsymbol{\Delta} = \begin{matrix} A \\ B \end{matrix} \begin{bmatrix} \ddots & & & & & & \\ \dots & -\Delta_2 s_\perp & -\Delta_1 + i\Delta'_1 s_\parallel & 0 & -\Delta_1 - i\Delta'_1 s_\parallel & \Delta_2 s_\perp & \dots \\ \dots & -\Delta_2 s_\perp^* & \Delta_1 + i\Delta'_1 s_\parallel & 0 & \Delta_1 - i\Delta'_1 s_\parallel & \Delta_2 s_\perp^* & \dots \\ & & & & & & \ddots \end{bmatrix}, \quad (6)$$

where $s_\parallel = \sin\vartheta \sin\varphi$ and $s_\perp = -\cos\varphi - i\cos\vartheta \sin\varphi$ are the components of the impurity spin parallel and perpendicular to the $y$ direction preferred by the SOC. The Hamiltonian of the antiferromagnetic chain possesses the ETRS $T_{\text{eff}} = U_B R_1 K$, where $R_1$ is translation by a lattice constant and $U_B$ adds a negative sign on sublattice $B$ and acts as identity on sublattice $A$.

For $s_\perp = 0$, the ETRS may be rewritten in Fourier space as a Kramers symmetry $T_{\text{eff}}^2 = -1$, enforcing the pairwise degeneracy of the state. If the impurity spins have a component perpendicular to the direction selected by the SOC, then one obtains $T_{\text{eff}}^2 \neq -1$ and the Kramers degeneracy or the ETRS is broken in this sense. This can be assumed to be the case here, since the spins have an out-of-plane component while the SOC selects an in-plane direction. If we set $s_\parallel = 0$ which does not influence whether ETRS is broken or not, and chose $s_\perp = 1$ to be real which can be achieved by an appropriate rotation of the spin quantization axes around $y$, then the local gauge transformation $c_j = (-1)^j U_B i \tilde{c}_j$ and $c_j^\dagger = (-1)^j U_B i \tilde{c}_j^\dagger$, where $(-1)^j U_B$ describes a sign change after every two lattice sites, transforms the Hamiltonian in Eqs. (2), (5) and (6) to the form given in Eq. (1) in the main text.

The following analysis is introduced in Ref. [31]. Here, we repeat the necessary details and adjust it to the antiferromagnetic system. Starting from Eq. (1) in the main text, the LDOS as a function of energy $E$ and position $x$ along a one-dimensional lattice of $N$ sites in Fig. 4d is computed by exact diagonalization of the low-energy Hamiltonian in Eq. (1) and summing over all pairs of eigenvalues $E_i$ and eigenvectors $\psi_i$:

$$\text{LDOS}(E, x) = \sum_i \left[ P |\psi_{i,e}(x)|^2 + (1-P)|\psi_{i,h}(x)|^2 \right] \left( -\frac{\partial f(E - E_i, T = 320 \text{ mK})}{\partial E} \right) \quad (7)$$

with the respective particle- (e) and hole-components (h) of the solutions and the Fermi-Dirac function $f(E, T)$ simulating the experimental thermal broadening. Here, $P = 0.2$ is chosen to account for the particle-hole asymmetric spectral weight observed when tunneling into YSR bands.

The presence of the ETRS along with the particle-hole constraint (symmetry class DIII for $T_{\text{eff}}^2 = -1$ and BDI for $T_{\text{eff}}^2 = 1$) suggests that the system should be described by a different topological invariant than the Majorana number defined for the Kitaev chain with only particle-hole constraint (symmetry class D)[10]. However, the bulk-boundary correspondence cannot be used to conclude on the presence of topologically protected edge modes based on a different classification, because the finite chain typically does not possess an ETRS. The translation along the chain described for the infinite chain above is obviously broken for a finite chain. Odd-length chains have a net magnetic moment; therefore, they cannot be described by any ETRS. Instead of the translation, another lattice symmetry could be combined with the physical time reversal to obtain $T_{\text{eff}}$ for an even-length chain. However, the mirror symmetries proposed in Ref. [10] do not hold for the chain built along the [1$\bar{1}$1] direction. A 180° rotation around the out-of-plane direction at the middle of the chain holds for the considered system, but perturbing one end of the chain breaks the 180° rotation symmetry. Based on the above, only the Majorana number $\mathcal{M}$ introduced for the Kitaev chain can be used as an indication for the presence of topologically non-trivial edge states, since this does not rely on the ETRS. This topological invariant $\mathcal{M}$ can be calculated as:

$$\mathcal{M} = \text{sgn}\{\text{Pf}[\widetilde{H}(0)]\text{Pf}[\widetilde{H}(\pi/d)]\} \qquad (8)$$

where Pf denotes the Pfaffian and $\widetilde{H}(k)$ is the *k*-space Hamiltonian in the Majorana basis[48]. For the present system $\text{Pf}[\widetilde{H}(k)] = -E_0 - t_k$, where $t_k = 2t_1 \cos kd + 2t_2 \cos 2kd$ is the Fourier transform of the hopping terms. The topological invariant $\mathcal{M}$ takes the value -1 (+1) for the YSR band crossing the Fermi level an odd (even) number of times between 0 and $\pi/d$ when the superconducting pairing terms are set to zero. The system is always in the topologically trivial regime in the antiferromagnetic case without SOC ($t_1 = 0$), where $t_k$ is $\pi/d$ periodic and the number of band crossings is even. Changing the parity of the band crossings requires adding strong ETRS-breaking terms to the Hamiltonian. For the extended model described by Eqs. (2), (5) and (6), the boundary of the Brillouin zone is reduced to $\pi/(2d)$ due to the antiferromagnetic structure, and there are two particle-hole pairs for each wave vector. In this case, $\mathcal{M}$ is given by the sign of the product of $\text{Pf}[\widetilde{H}(0)] = (-E_0 + 2t_2)^2 + (2\Delta_1)^2$ and $\text{Pf}[\widetilde{H}(\pi/(2d))] = (-E_0 - 2t_2)^2 + (2\Delta'_1 s_\parallel)^2 - |2h_1 s_\perp|^2$, which simplifies to the same condition as above for $s_\parallel = 0$ and $s_\perp = 1$, and also demonstrates that a finite value of $s_\parallel$ does not prefer the formation of a topologically non-trivial state.

## References

51. Pientka, F., Glazman, L. I. & von Oppen, F. Unconventional topological phase transitions in helical Shiba chains. *Phys. Rev. B* **89**, 180505 (2014).
52. Wiebe, J. *et al.* A 300 mK ultra-high vacuum scanning tunneling microscope for spin-resolved spectroscopy at high energy resolution. *Rev. Sci. Instrum.* **75**, 4871–4879 (2004).
53. Odobesko, A. B. *et al.* Preparation and electronic properties of clean superconducting Nb(110) surfaces. *Phys. Rev. B* **99**, 115437 (2019).
54. Pan, S. H., Hudson, E. W. & Davis, J. C. Vacuum tunneling of superconducting quasiparticles from atomically sharp scanning tunneling microscope tips. *Appl. Phys. Lett.* **73**, 2992–2994 (1998).
55. von Oppen, F., Peng, Y. & Pientka, F. *Topological superconducting phases in one dimension: Lecture Notes of the Les Houches Summer School*. *Topological Aspects of Condensed Matter Physics* (Oxford Univ. Press, 2014).

**Data availability**
The authors declare that the data supporting the findings of this study are available within the paper and its supplementary information files.

**Code availability**
The analysis codes that support the findings of the study are available from the corresponding authors on reasonable request.

## Acknowledgements
We thank Eric Mascot, Roberto Lo Conte and Falko Pientka for helpful discussions. L.S., T.P., J.W., and R.W. gratefully acknowledge funding by the Cluster of Excellence 'Advanced Imaging of Matter' (EXC 2056 - project ID 390715994) of the Deutsche Forschungsgemeinschaft (DFG). L.R. gratefully acknowledges financial support from the National Research, Development and Innovation Office of Hungary via Project No. K131938 and from the Young Scholar Fund at the University of Konstanz. P.B., J.W. and R.W. acknowledge support by the DFG (SFB 925 – project 170620586). R.W. acknowledges funding by the European Union via the ERC Advanced Grant ADMIRE (project No. 786020). T.P. acknowledges support by the DFG (project no. 420120155).

**Author contributions**
L.S., P.B., R.W. and J.W. conceived the experiments. L.S. and P.B. performed the measurements and analyzed the experimental data. L.S. performed the numerical simulations in close exchange with T.P. and L.R.. L.S. prepared the figures and wrote the manuscript. All authors contributed to the discussions and to correcting the manuscript.

**Competing interests**
The authors declare no competing interests.

# Testing the topological nature of end states in antiferromagnetic atomic chains on superconductors

**Supplementary Information**

Lucas Schneider, Philip Beck, Levente Rózsa, Thore Posske, Jens Wiebe and Roland Wiesendanger

Correspondence to: jwiebe@physnet.uni-hamburg.de

## Supplementary Note 1 | Calculations on the stability of states against local disorder

In Fig. 3 of the main manuscript text, the response of end states to local potential changes is studied. In the model introduced in Fig. 4 and in the Methods section of the main manuscript text, this type of disorder would correspond to disorder in the YSR energy $E_0$ at individual sites of the chain. Depending on the type of state that is perturbed by disorder, a different response is expected. In particular, the more local a certain state is, the more it is expected to react to disorder. To sketch this effect, we use the model presented in the Methods section of the main manuscript text and add randomly distributed values from the interval $[-\delta E_0, \delta E_0]$ *to* the on-site YSR energies $E_{0,i}$ for every site $i$ of the chain. The results are shown in Supplementary Fig. 1: for states perfectly localized on one individual site (Supplementary Fig. 1a), the spread in eigenenergies is – as expected – simply linearly increasing with the magnitude of the noise in the potential with slope of 1. In contrast, MMs are expected to be protected against disorder due to their non-local nature[1]. Supplementary Fig. 1b shows the same evolution of the eigenvalues with increasing disorder for a chain with *N* = 21 sites in the small-gap topologically non-trivial regime where the precursors of MMs (PMMs) still oscillate strongly in energy with increasing chain length due to large overlap of the Majorana wave functions[2]. Notably, the near-zero energy mode is much more stable against local disorder than the atomic states in Supplementary Fig. 1a. However, this protection is not much stronger than the one from the finite-size quantized states[3,4] at finite energies, which also split only weakly for moderate disorder strength. This is a result of the spatially extended nature of both PMMs and the finite-size quantized states. In contrast, the MMs in Supplementary Fig. 1c, showing the large-gap scenario of the topologically non-trivial phase, are much more protected compared to all other states and remain at zero energy even for large values of $\delta E_0$ because of their clear nonlocality. Although the magnetic structure of the chain does not enter the minimal model explicitly, the parameter values chosen in Supplementary Fig. 1b and c can be thought of representing a ferromagnetic chain where the minigap is opened by the spin-orbit coupling. Finally, the phase with $t_2 \gg t_1$ analyzed in Fig. 4 of the main manuscript text, modelling an antiferromagnetic chain, is studied in Supplementary Fig. 1d. The end states at finite energy of $E \approx \pm 0.5$ meV react strongly to disorder. This is highlighted by the blue dashed lines in Supplementary Fig. 1d, indicating the (maximal) splitting of the states in Supplementary Fig. 1a, showing that the response of the end states is about 50% as strong as for the uncoupled sites. The effect can be understood to be a consequence of the strongly localized nature of the end states. Overall, these theoretical results agree with the experimental findings of Fig. 3 in the main manuscript text where the end states were found to split by several tens of µeV when perturbing them with magnetic or non-magnetic defects. It should be noted that they also split into two particle-hole pairs (i.e. four solutions in total) for every distribution of $\delta E_0$ whereas MMs and their precursors always split into a single particle-hole pair. This further supports the interpretation that our model maps well onto the experimental platform.

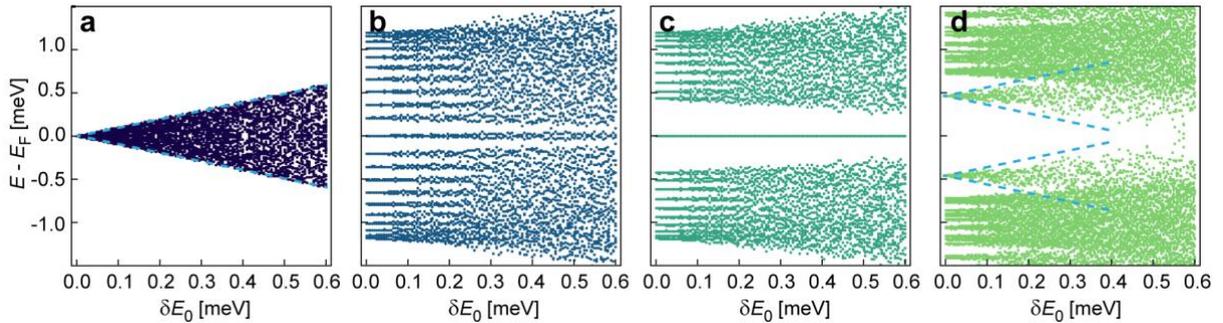

**Supplementary Figure 1 | Stability of end states against local potential noise. a**, Eigenstates of a chain of *N* = 21 uncoupled sites ($t_1 = t_2 = \Delta_1 = \Delta_2 = 0.0$ meV) with increasing potential noise $\delta E_0$ added to the on-site potential $E_0 = 0$ meV. Obviously, the distribution width of the eigenstates spreads linearly with a slope of 1 (blue dashed lines). **b**, Eigenstates of a chain of *N* = 21 sites in the small-gap topologically non-trivial regime vs. increasing potential noise $\delta E_0$. Parameters: $t_1 = 0.6$ meV, $t_2 = 0.0$ meV, $\Delta_1 = 0.06$ meV, $\Delta_2 = 0.0$ meV. **c**, Eigenstates of a chain of *N* = 21 sites in the large-gap topologically non-trivial regime vs. increasing potential noise $\delta E_0$. Parameters: $t_1 = 0.6$ meV, $t_2 = 0.0$ meV, $\Delta_1 = 0.2$ meV, $\Delta_2 = 0.0$ meV. **d**, Eigenstates of a chain of *N* = 21 sites with trivial end states in the regime $t_2 \gg t_1$ vs. increasing potential noise $\delta E_0$. Blue dashed lines represent the slope of 1 as also shown in panel a. Parameters: $t_1 = 0.0$ meV, $t_2 = 0.6$ meV, $\Delta_1 = 0.5$ meV, $\Delta_2 = 0.0$ meV.

## Supplementary Note 2 | Spin-polarized measurements on Mn/Ta(110)

The antiferromagnetic ground state of densely packed Mn chains along the [1$\bar{1}$1] direction on Nb(110) was determined in Ref. [5] and confirmed by *ab-initio* calculations in Ref. [6]. Yet, the magnetic couplings of Mn adatoms on Ta(110) are not *a priori* known. Therefore, we performed additional measurements with spin-polarized Cr tips[5,7] on Mn chains along [1$\bar{1}$1] on Ta(110). Cr tips were made from high-purity Cr splinters glued into a W tip holder with conductive H20E glue[8]. The tip was subsequently heated to $T \approx 700$ K *in situ* and voltage pulses of 10 V were applied against a Pt(111) surface in order to remove oxide layers from the tip apex. Supplementary Fig. 2a shows an STM topography image of a $Mn_9$ chain on Ta(110) imaged with a Cr tip in a weak external field of $B_z = +20$ mT. Four maxima are observed on the chain, corresponding to the spatial positions of the atom No. 2, 4, 6 and 8. When reversing the field direction ($B_z = -20$ mT), imaging the same chain results in a pattern with three maxima on atoms No. 3, 5 and 7 and minima in between. These two images can be interpreted as follows: the Mn chain is antiferromagnetically aligned, yielding an alternating contrast on neighboring sites. While the Cr tip's magnetization is stable in an external field, the degeneracy of the two collinear antiferromagnetic Néel states sketched above the panels is slightly lifted by the external field acting on a structure with an odd number of sites[9]. The terminal chain sites always feature an electronic contrast which makes it hard to interpret the spin-contrast on them. This could be the reason why 3 and 4 maxima are found in the images instead of 4 and 5. Note that the spin-contrast is detected in the Z signal since we use very small bias voltages of $V_{stab}$ = 1 mV.

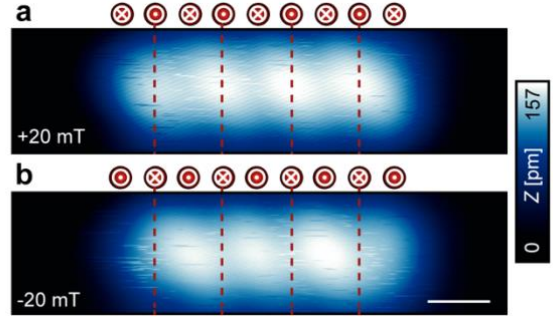

**Supplementary Figure 2 | Spin-polarized STM images of the magnetic ground state of Mn chains on Ta(110). a**, Constant-current STM image of a $Mn_9$ chain on Ta(110) in a weak out-of-plane magnetic field of $B_z = +20$ mT using a spin-polarized Cr tip with stable magnetization. The sketch above the image illustrates the antiferromagnetic spin texture of the nine atoms which are stabilized into one of the Néel ground states by the external field. **b**, Constant-current STM image of the same $Mn_9$ chain measured in a reversed field of $B_z = -20$ mT, revealing opposite spin contrast. Above the image, the reversed spins compared to panel a are sketched. Dashed lines are guides to the eye. The white scale bar corresponds to 500 pm. Parameters: $V_{stab}$ = 1 mV, $I_{stab}$ = 0.2 nA.

## Supplementary Note 3 | Determination of tip and sample gaps

The numerical deconvolution process performed on the data presented in the main figures (see Methods for details) requires sufficient knowledge of the tip density of states $\rho_t(E)$. We assume a broadened Dynes density of states for the superconducting Nb tip apex, given by:

$$\rho_t(E) = \rho_0 Re\left[\frac{E - i\Gamma}{\sqrt{(E - i\Gamma)^2 - \Delta_t^2}}\right]. \quad (S1)$$

Here, $\rho_0$ denotes the normal conducting DOS, $Re$ is the real part, $\Gamma$ is the Dynes broadening parameter and $\Delta_t$ is the tip's superconducting gap. Throughout this work, $\Gamma$ = 0.01 meV is chosen for all tips used. The value of $\Delta_t$ is estimated by measurements of multiple Andreev reflections (MARs) appearing as weak additional peaks in d$I$/d$V$ spectra for low junction resistances[10]. This effect is shown in Supplementary Fig. 3 for tunneling between a superconducting Nb tip and a clean Ta(110) sample surface: for high junction resistances (Supplementary Fig. 3a), the convolution of the sample's and tip's coherence peaks yields a large peak in d$I$/d$V$ at $eV = \pm(\Delta_t + \Delta_{Ta})$, where $\Delta_{Ta}$ is the gap of the Ta sample.

In contrast, for lower junction resistances (Supplementary Fig. 3b), additional peaks at $\pm 1.25$ mV and $\pm 0.64$ mV are found. Since the latter value matches the superconducting gap of Ta while the first value is only slightly smaller than the gap of Nb, we attribute these peaks to Andreev tunneling occurring at $eV = \pm\Delta_t$ and $eV = \pm\Delta_{Ta}$, respectively[10]. Importantly, the MAR peaks are not visible in the high resistance limit (Supplementary Fig. 3a), indicating that single-particle tunneling is the dominant transport channel and that the deconvolution process is justified. The result of a numerical deconvolution of the spectrum shown in Supplementary Fig. 3a is presented in Supplementary Fig. 3c. Indeed, this spectrum is well described by a Dynes density of states (see Eq. S1) as well when choosing a broadening parameter of $\Gamma$ = 0.03 meV and a gap $\Delta_{Ta}$ = 0.64 meV.

The same characterization for measurements on the clean Nb(110) surface can be found in the Supplementary Information of Ref. [3].

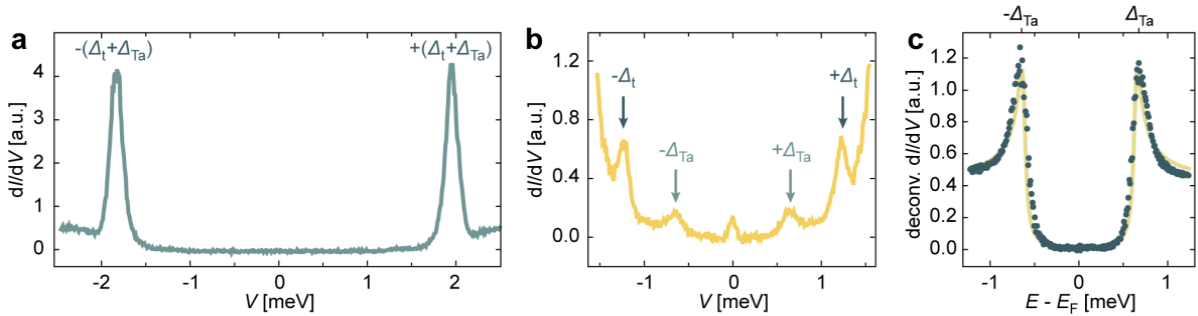

**Supplementary Figure 3 | Determination of tip and sample gaps for measurements on Ta(110). a**, d$I$/d$V$ spectrum measured on the clean Ta(110) surface at high junction resistance ($V_{stab}$ = -2.5 mV, $I_{stab}$ = 1 nA, $V_{mod}$ = 20 μV), showing prominent peaks at bias voltages $e \cdot V = \pm(\Delta_t + \Delta_{Ta})$ and no additional sub-gap peaks. **b**, d$I$/d$V$ spectrum measured with the same tip at low junction resistance ($V_{stab}$ = -2.5 mV, $I_{stab}$ = 10 nA, $V_{mod}$ = 20 μV) with distinct additional peaks visible at $e \cdot V = \pm\Delta_t = 1.25$ meV and $e \cdot V = \pm\Delta_{Ta}$ due to multiple Andreev reflection processes. Furthermore, Josephson tunneling occurs at zero voltage, yielding a zero-bias peak. **c**, Spectrum from panel a after numerical deconvolution (see Methods) using the previously determined tip gap parameter $\Delta_t$ from panel b. Assuming a superconducting density of states modeled by the Dynes function with $\Gamma = 0.03\ meV$ and $\Delta_{Ta} = 0.64\ meV$ (solid line) describes the data (dark points) well.

## Supplementary Note 4 | Sub-gap quasiparticle interference measurements

The standing-wave-like patterns observed experimentally in Figs. 1b,d of the main manuscript can be explained by interference of energetically degenerate sub-gap quasiparticles with momenta $k_i$ and $k_f$ in the YSR bands[2,3]. As sketched in Fig 4a of the main manuscript, two dominant QPI branches with scattering vectors $q_1$ and $q_2$ ($|q| = |k_i - k_f|$) are expected based on the minimal model of the main manuscript text for small $E_0$ (i.e. deep YSR states), $t_1 \ll t_2$ and $\Delta_1 \gg \Delta_2$ (appropriate for an antiferromagnetic chain). By performing a line-wise fast Fourier transform (FFT) analysis of the experimental data in Fig. 1d of the main manuscript, information about the YSR band dispersion can be extracted (Fig. 4c). Note that the FFT analysis does not directly display the band structure but the possible scattering vectors $q$ in the QPI process.

Focusing on the hole-like part of the data in Fig. 4c ($E > 0$), two dominant arcs around $q = 0$ are observed, one with negative (supposedly $q_1$) and one with positive (supposedly $q_2$) curvature. Comparing these to the predicted scattering vectors in Fig. 4a and the theoretically simulated QPI pattern in Fig. 4b, it is found that $q_1$ is expected to be close to $q/2 = \pm\pi/4$ at the bottom of the hole-like band (indicated by the gray dashed lines), which is the case for the supposed $q_1$ branch extracted from the experimental data (Fig. 4c). Additionally, QPI from scattering of type $q_2$ (Fig. 4a) would result in a branch with positive effective mass, which could be the second arc marked in Fig. 4c. The electron-like part of the figure can be explained in the same way, although asymmetries in the particle- and hole-spectral-weight of the individual YSR states[11,12] can lead to a slightly different appearance of the QPI patterns[3]. Interestingly, the end states at $\varepsilon_{+/-} \approx \pm 0.5$ meV also feature a prominent Fourier component of $q/2 = \pm\pi/4$. Additional frequency components, e.g. at $\pm\pi/8$ or zero, can be attributed to artifacts of the FFT acting on a peaked signal like an end state. This suggests that the end states inherit properties like the wavelength from the observed dispersive sub-gap band. A similar effect is known for topological MMs[13,14].

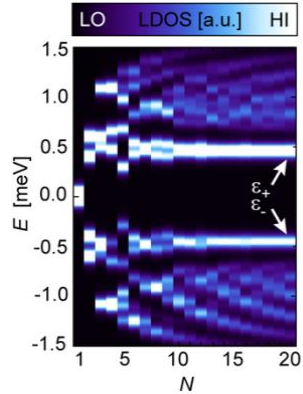

**Supplementary Figure 4 | Calculated evolution of the end states with the length of the antiferromagnetic chains.** Evolution of the LDOS calculated using the minimal model described in the main manuscript text on the terminal two sites of a chain with increasing number of sites $N$. The used parameters are $E_0 = 0.0$ meV, $t_1 = 0.1$ meV, $t_2 = 0.6$ meV, $\Delta_1 = 0.5$ meV, and $\Delta_2 = 0.0$ meV. For $N$ = 1 and 2, there are no terminal sites and the total DOS is shown. The energies of the finite energy end states $\varepsilon_{+/-}$ are marked.

# References

1. Awoga, O. A., Björnson, K. & Black-Schaffer, A. M. Disorder robustness and protection of Majorana bound states in ferromagnetic chains on conventional superconductors. *Phys. Rev. B* **95**, 184511 (2017).
2. Schneider, L. *et al.* Precursors of Majorana modes and their length-dependent energy oscillations probed at both ends of atomic Shiba chains. *Nat. Nanotechnol.* **17**, 384–389 (2022).
3. Schneider, L. *et al.* Topological Shiba bands in artificial spin chains on superconductors. *Nat. Phys.* **17**, 943–948 (2021).
4. Ben-Shach, G. *et al.* Detecting Majorana modes in one-dimensional wires by charge sensing. *Phys. Rev. B* **91**, 045403 (2015).
5. Schneider, L., Beck, P., Wiebe, J. & Wiesendanger, R. Atomic-scale spin-polarization maps using functionalized superconducting probes. *Sci. Adv.* **7**, eabd7302 (2021).
6. Lászlóffy, A., Palotás, K., Rózsa, L. & Szunyogh, L. Electronic and Magnetic Properties of Building Blocks of Mn and Fe Atomic Chains on Nb(110). *Nanomaterials* **11**, 1933 (2021).
7. Schlenhoff, A., Krause, S., Herzog, G. & Wiesendanger, R. Bulk Cr tips with full spatial magnetic sensitivity for spin-polarized scanning tunneling microscopy. *Appl. Phys. Lett.* **97**, 083104 (2010).
8. EPOXY TECHNOLOGY, INC. - 14 Fortune Drive, Billerica, Massachusetts 01821, USA.
9. Khajetoorians, A. A. *et al.* Atom-by-atom engineering and magnetometry of tailored nanomagnets. *Nat. Phys.* **8**, 497–503 (2012).
10. Ternes, M. *et al.* Subgap structure in asymmetric superconducting tunnel junctions. *Phys. Rev. B* **74**, 132501 (2006).
11. Balatsky, A. V., Vekhter, I. & Zhu, J.-X. Impurity-induced states in conventional and unconventional superconductors. *Rev. Mod. Phys.* **78**, 373–433 (2006).
12. Ruby, M. *et al.* Tunneling Processes into Localized Subgap States in Superconductors. *Phys. Rev. Lett.* **115**, 087001 (2015).
13. Klinovaja, J. & Loss, D. Composite Majorana fermion wave functions in nanowires. *Phys. Rev. B* **86**, 085408 (2012).
14. Peng, Y., Pientka, F., Glazman, L. I. & von Oppen, F. Strong Localization of Majorana End States in Chains of Magnetic Adatoms. *Phys. Rev. Lett.* **114**, 106801 (2015).